\documentclass[fleqn,usenatbib]{mnras}

\usepackage{newtxtext,newtxmath}

\usepackage[T1]{fontenc}
\usepackage{ae,aecompl}
\usepackage{hyperref}
\newcommand{\printfnsymbol}[1]{%
  \textsuperscript{\@fnsymbol{#1}}%
}

\usepackage{graphicx}	
\usepackage{amsmath}	
\usepackage{amssymb}	

\title[3-body problem and deep neural networks]{Newton vs the machine: solving the chaotic three-body problem using deep neural networks}
\author[Breen, Foley, Boekholt,\&  Portegies Zwart]{
Philip G. Breen$^{1}$\thanks{Authors contributed equally}\thanks{Contact e-mail: \href{mailto:phil.breen@ed.ac.uk}{phil.breen@ed.ac.uk}},
Christopher N. Foley$^{2}$ \footnotemark[1]\thanks{Contact e-mail: \href{mailto:christopher.foley@mrc-bsu.cam.ac.uk}{christopher.foley@mrc-bsu.cam.ac.uk}},
Tjarda Boekholt$^{3}$\newauthor{\,\,and Simon Portegies Zwart$^{4}$}
\\
$^{1}$School of Mathematics and Maxwell Institute for Mathematical Sciences, University of Edinburgh, Kings Buildings, Edinburgh, EH9 3JZ\\
$^2$MRC Biostatistics Unit, University of Cambridge, Cambridge, CB2 0SR, UK.\\ 
$^3$Instituto de Telecomunica\c{c}\~oes, Campus Universit\'ario de Santiago, 3810-193, Aveiro, Portugal\\
$^4$Leiden Observatory, Leiden University, PO Box 9513, 2300 RA, Leiden, The Netherlands.\\
}

\date{Accepted XXX. Received YYY; in original form ZZZ}

\pubyear{2019}

\begin{document}
\label{firstpage}
\pagerange{\pageref{firstpage}--\pageref{lastpage}}
\maketitle

\begin{abstract}
Since its formulation by Sir Isaac Newton, the problem of solving the equations of motion for three bodies under their own gravitational force has remained practically unsolved. Currently, the solution for a given initialization can only be found by performing laborious iterative calculations that have unpredictable and potentially infinite computational cost, due to the system's chaotic nature. We show that an ensemble of solutions obtained using an arbitrarily precise numerical integrator can be used to train a deep artificial neural network (ANN) that, over a bounded time interval, provides accurate solutions at fixed computational cost and up to 100 million times faster than a state-of-the-art solver. Our results provide evidence that, for computationally challenging regions of phase-space, a trained ANN can replace existing numerical solvers, enabling fast and scalable simulations of many-body systems to shed light on outstanding phenomena such as the formation of black-hole binary systems or the origin of the core collapse in dense star clusters.
\end{abstract}




\section{Introduction}

Newton's equations of motion describe the evolution of many bodies in space under the influence of their own gravitational force \citep{Newton1687}. The equations have a central role in many classical problems in Physics. For example, the equations explain the dynamical evolution of globular star clusters and galactic nuclei, which are thought to be the production sites of tight black-hole binaries that ultimately merge to produce gravitational waves \citep{PZMM2000}. The fate of these systems depends crucially on the three-body interactions between black-hole binaries and single black-holes \citep[e.g. see][]{BH2013A,BH2013B,SD2018}, often referred to as  close encounters. These events typically occur over a fixed time interval and, owing to the tight interactions between the three nearby bodies, the background influence of the other bodies can be ignored, i.e. the trajectories of three bodies can be generally computed in isolation \citep{PZMM2018}. The focus of the present study is therefore the timely computation of accurate solutions to the three-body problem.

Despite its age and interest from numerous distinguished scientists \citep{dL1772,H1975,HB1983,M1998,SL2019}, the problem of solving the equations of motion for three-bodies remains impenetrable due to the system's chaotic nature \citep{V2016} which typically renders identification of solutions feasible only through laborious numerical integration. Analytic solutions exist for several special cases \citep{dL1772} and a solution to the problem for all time has been proposed \citep{V2016}, but this is based on an infinite series expansion and has limited use in practice. Computation of a numerical solution, however, can require holding an exponentially growing number of decimal places in memory and using a time-step that approaches zero \citep{BPZV2019}. Integrators which do not allow for this often fail spectacularly, meaning that a single numerical solution is unreliable whereas the average of an ensemble of numerical solutions appear valid in a statistical sense, a concept referred to as nagh Hoch \citep{PZB2018}. To overcome these issues, the Brutus integrator was developed \citep{BPZ2015}, allowing for close-to-zero time-steps and arbitrary precision. Brutus is capable of computing converged solutions to any gravitational N-body problem, however the process is laborious and can be extremely prohibitive in terms of computer time. In general, there does not exist a theoretical framework capable of determining a priori the precision required to deduce that a numerical solution has converged for an arbitrary initialization \citep{SL2019}. This makes the expense of acquiring a converged solution through brute-force integration unpredictable and regularly impractical. 

Here we demonstrate that, over a fixed time interval, the 300-year-old three-body problem can be solved by means of a multi-layered deep artificial neural network \citep[ANN, e.g. see][]{LBH2015}. These networks are designed for high-quality pattern recognition by mirroring the function of our brains \citep{MCP1943, R1958} and have been successfully applied to a wide variety of pattern recognition problems in science and industry, even mastering the game of Go \citep{Sat2016}.  The abundance of real-world applications of ANNs is largely a consequence of two properties: (i) an ANN is capable of closely approximating any continuous function that describes the relationship between an outcome and a set of covariates, known as the universal approximation theorem \citep{H1991,C1989}, and; (ii) once trained, an ANN has a predictable and a fixed computational burden. Together, these properties lead to the result that an ANN can be trained to provide accurate and practical solutions to Newton's laws of motion, resulting in major improvements in computational economy  \citep{LSYP1998} relative to modern technologies. Moreover, our proof-of-principle method shows that a trained ANN can accurately match the results of the arbitrary precision numerical integrator which, for computationally challenging scenarios, e.g. during multiple close encounters, can offer numerical solutions at a fraction of the time cost and CO2 expense. Our findings add to the growing body of literature which supports machine learning technologies being developed to enrich the assessment of chaotic systems \citep{Pat2018,S2019} and providing alternative approaches to classical numerical solvers more broadly \citep{HOG2015}.



\section{Method}

Every ANN requires a learning phase, where parameters in an adaptive model are tuned using a training dataset, which renders prediction accuracy sensitive to whether the training set is representative of the types of patterns that are likely present in future data applications.  Training an ANN on a chaotic problem therefore requires an ensemble of solutions across a variety of initializations. The only way to acquire such a training set is by numerically integrating the equations of motion for a large and diverse range of realizations until a converged solution is acquired, which we do using Brutus (an arbitrary precise N-body numerical integrator).   

\begin{figure}
    \centering
    \includegraphics[width=0.5\textwidth]{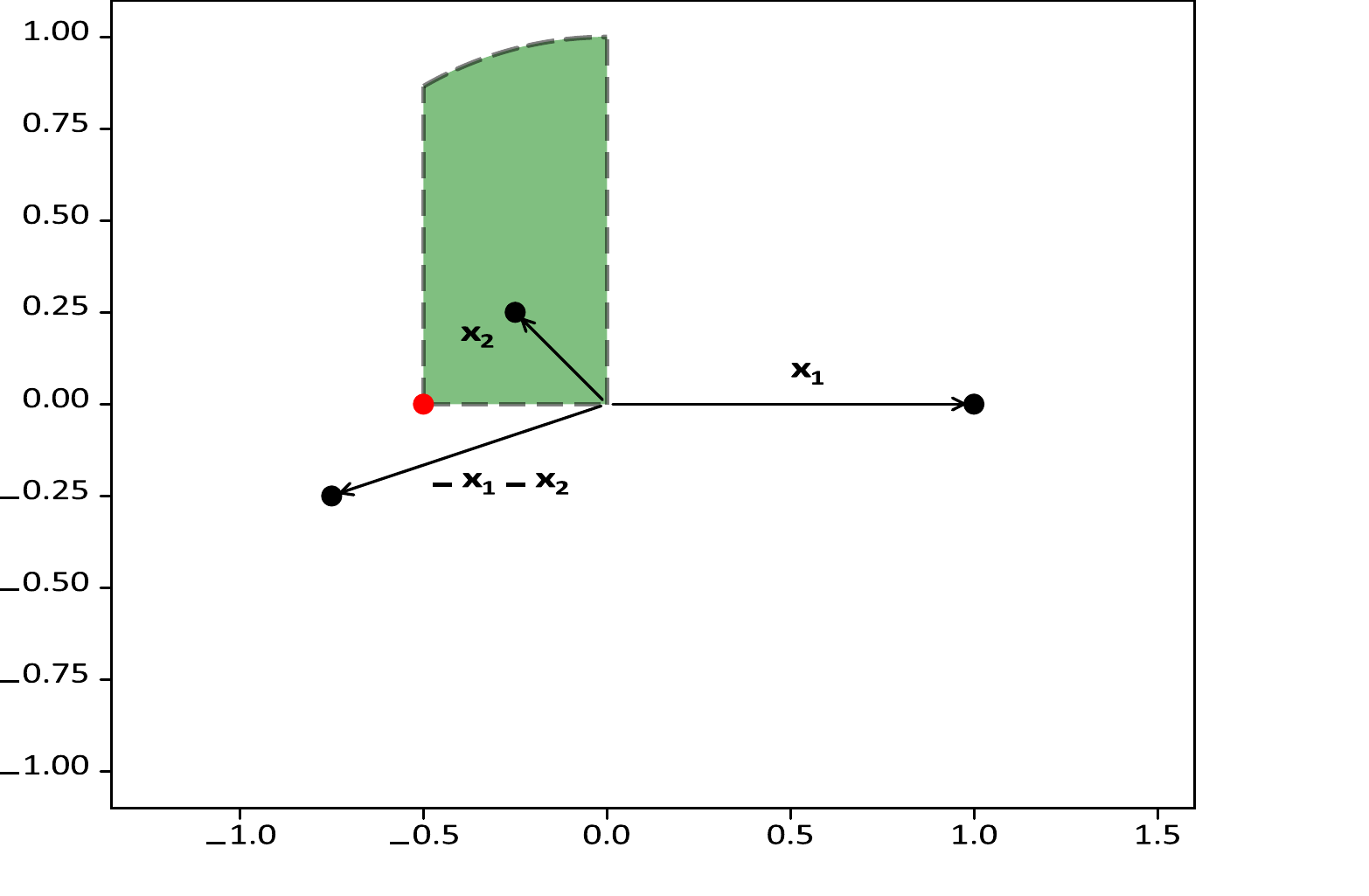}
    \caption{{\bf Visualization of the initial particle locations.} The origin is taken as the barycenter and the unit of length is chosen to be the distance to the most distant particle $x_1$, which also orientates the x-axis. The closest particle to the barycenter, labelled $x_2$, is chosen to orientate the positive y-axis and can be located anywhere in the green region. Once specified, the location of the remaining particle $x_3$ is deduced by symmetry. There is a singular point at (-0.5,0), red point, where the position of $x_2$ and $x_3$ are identical. Numerical schemes can fail near the point as the particles are on near collision orbits, i.e. passing arbitrarily close to one another.}
    \label{fig:initialcond}
\end{figure}

We restricted the training set to the gravitational problem of three equal mass particles with zero initial velocity, located in a plane.  The three particles, with Cartesian co-ordinates $x_1$, $x_2$, $x_3$, are initially positioned at $x_1 \equiv (1,0)$  with ($x_2$,$x_3$) randomly situated somewhere in the unit semicircle in the negative x-axis, i.e. $x \leq 0$. The reference frame is taken as the centre of mass and, without loss of generality, we orientate the positive y-axis using the particle closest to the barycentre (Fig. \ref{fig:initialcond}).  In this system, the initial location of only one of ($x_2$, $x_3$) need be specified, as the location of the remaining particle is deduced by symmetry. In addition, we adopt dimensionless units in which $G=1$ \citep{HM1986}. The physical setup allows the initial conditions to be described by 2 parameters and the evolution of the system by 3 parameters (representing the coordinates of $x_1$ and $x_2$ at a given time). The general solution is found by mapping the 3-dimensional phase-space (time $t$ and initial coordinate of $x_2$) to the positions of particles $x_1$ and $x_2$, the position of particle $x_3$ follows from symmetry.

The training and validation datasets are composed of 9900 and 100 simulations respectively. In each simulation, we randomly generated initial locations for the particles and computed trajectories, typically for up to 10 time-units (of roughly a dynamical crossing time each), by integrating the equations of motion using Brutus. Each trajectory comprises a dataset of some 2561 discrete time-points (labels), hence the validation dataset contained over $10^5$ time-points. A converged solution was acquired by iteratively reducing two parameters during integration: (i) the tolerance parameter ($\epsilon$), controlling accuracy, that accepts convergence of the Bulirsch-Stoer multi-step integration scheme \citep{BS1964} and; (ii) the word length ($L_w$) measured in bits, which controls numerical precision \citep{BPZ2015}.  Our ensemble of initial realizations all converged for values of $\epsilon = 10^{-11}$ and $L_w=128$ (see Appendix \ref{app:Brutus}). Generating these data required over 10 days of computer time. Some initialisations gave rise to very close encounters between the particles, e.g. mirroring a close encounters, and computation of converged solutions in these situations is costly\footnote{We note that identifying converged solutions for initial conditions near the singular point (0.5, 0) proved challenging. They result in very close encounters between two particles which could not be resolved within the predetermined precision. Brutus could have resolved these trajectories with higher precision, however this could result in even more lengthy computation time.} \citep{BPZV2019}.

\begin{figure}
    \centering
    \includegraphics[width=0.5\textwidth]{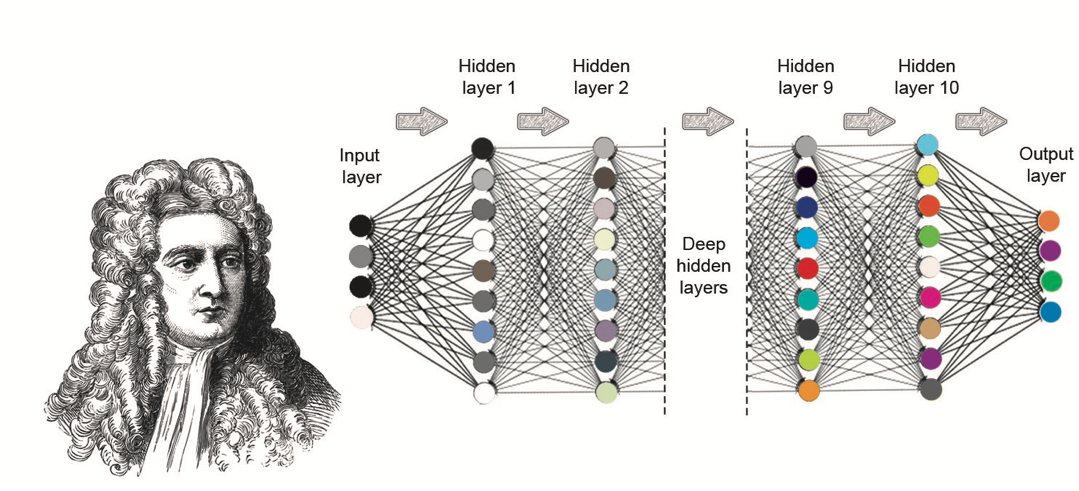}
    \caption{{\bf Newton and the machine.}  Image of sir Isaac Newton alongside a schematic of a 10-layer deep neural network. In each layer (apart from the input layer), a node takes the weighted input from the previous layer's nodes (plus a bias) and then applies an activation function before passing data to the next node. The weights (and bias) are free parameters which are updated during training. }
    \label{fig:newtonann}
\end{figure}

We used a feed-forward ANN consisting of 10 hidden layers of 128 interconnected nodes (Fig. \ref{fig:newtonann} and Appendix \ref{app:DNN}). Training was performed using the adaptive moment estimation optimization algorithm ADAM (20) with 10000 passes over the data, in which each epoch was separated into batches of 5000, and setting the rectified linear unit (ReLU) activation function to $\max(0,x)$ \citep{GBB2011}. By entering a time t and the initial location of particle $x_2$ into the input layer, the ANN returns the locations of the particles $x_1$ and $x_2$ at time $t$, thereby approximating the latent analytical solution to the general three-body problem. 

\begin{figure}
    \centering
    \includegraphics[width=0.45\textwidth]{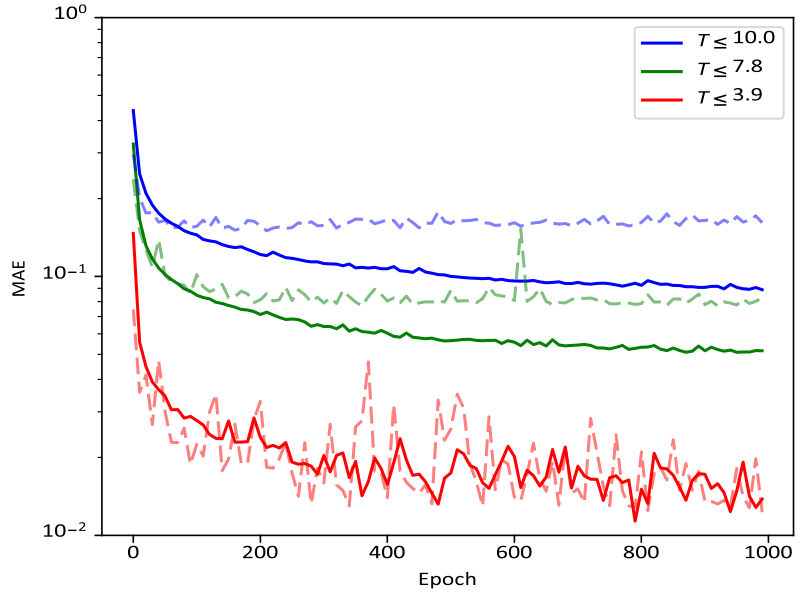}
    \caption{{\bf  Mean Absolute Error (MAE) vs epoch.}  The ANN has the same training structure in each time interval. Solids lines are the loss on the training set and dashed are the loss on the validation set. $T \leq 3.9$ corresponds to 1000 labels per simulation, similarly  $T \leq 7.8$  to 2000 labels and  $T \leq 10.0$ to 2561 labels/time-points (the entire dataset).The results illustrate a typical occurrence in ANN training, there is an initial phase of rapid learning, e.g. ≲100 epochs, followed by a stage of much slower learning in which relative prediction gains are smaller with each epoch. }
    \label{fig:MAE}
\end{figure}

To assess performance of the trained ANN across a range of time intervals, we partitioned the training and validation datasets into three segments: $t\lesssim 3.9$, $t \lesssim 7.8$ and $t \lesssim 10$
(which includes all data). For each scenario, we assessed the loss-function (taken as the mean absolute error MAE) against epoch. Examples are given in Fig. \ref{fig:MAE}. In all scenarios the loss in the validation set closely follows the loss in the training set. We also assessed sensitivity to the choice of activation function, however no appreciable improvement was obtained when using either the exponential rectified \citep{CUH2015} or leaky rectified \citep{MHN2013} linear unit functions. In addition, we assessed the performance of other optimization schemes for training the ANN, namely an adaptive gradient algorithm \citep{DHS2011} and a stochastic gradient descent method using Nesterov momentum, but these regularly failed to match the performance of the ADAM optimizer.

\begin{figure}
    \centering
    \includegraphics[width=0.45\textwidth]{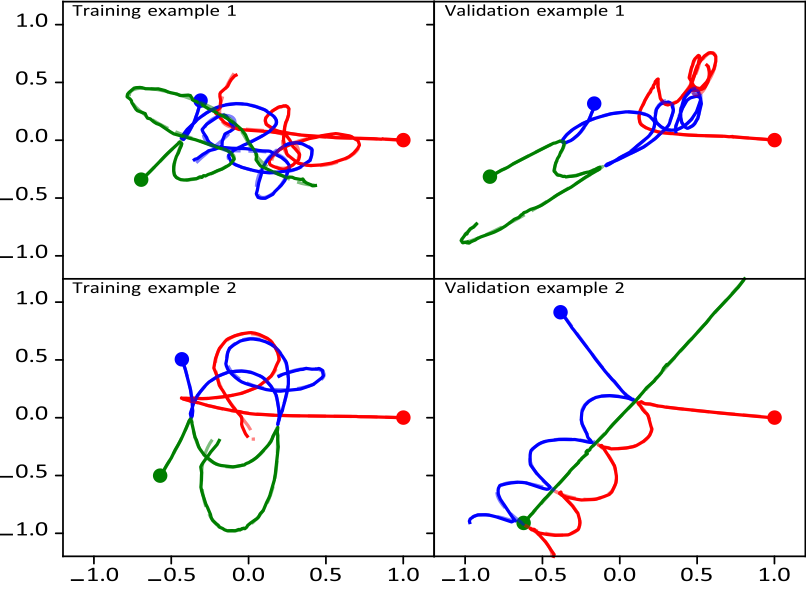}
    \caption{{\bf  Validation of the trained ANN.}  Presented are two examples from the training set (left) and two from the validation set (right).  All examples were randomly chosen from their datasets. The bullets indicate the initial conditions.  The curves represent the orbits of the three bodies (red, blue and green, the latter obtained from symmetry). The solution from the trained network (solid curves) is hardly distinguishable from the converged solutions (dashes, acquired using Brutus \citep{BPZ2015}).  The two scenarios presented to the right were not included in the training dataset. }
    \label{fig:traj}
\end{figure}

The best performing ANN was trained with data from $t \lesssim 3.9$ (Fig. \ref{fig:MAE}). We give examples of predictions made from this ANN against converged solutions within the training set (Fig. \ref{fig:traj}, left) or the validation set (Fig. \ref{fig:traj}, right). In each scenario, the particle trajectories reflect a series of complex interactions and the trained ANN reproduced these satisfactorily (MAE $\leq0.1$). The ANN also closely matched the complicated behaviour of the converged solutions in all the scenarios that were not included in its training. Moreover, the ANN did this in fixed computational time ($t \sim 10^{-3}$ seconds) which is on average about $10^5$ (and sometimes even $10^8$) times faster than Brutus. 

\begin{figure}
    \centering
    \includegraphics[width=0.45\textwidth]{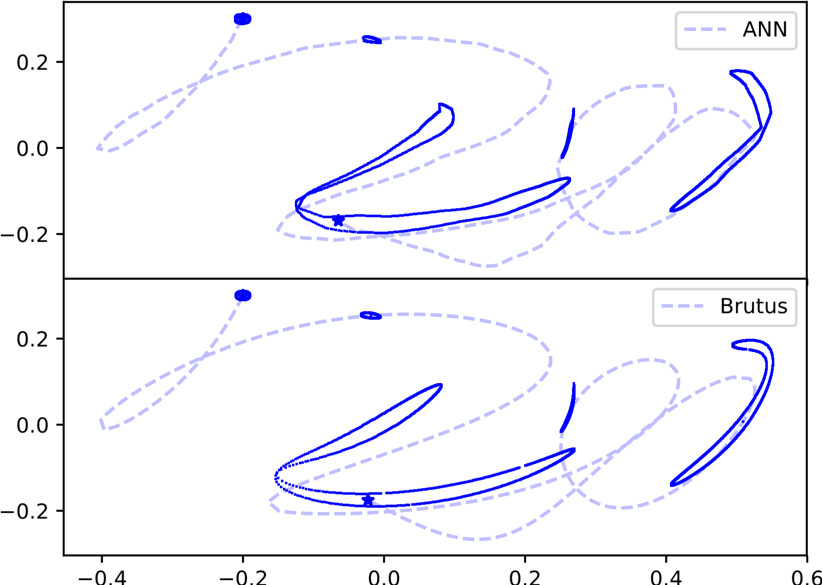}
    \caption{{\bf  Visualization of the sensitive dependence on initial position.} Presented are trajectories from 1000 random initializations in which particle $x_2$ is initially situated on the circumference of a ring of radius 0.01 centred at (-0.2, 0.3). For clarity, these locations were benchmarked against the trajectory of $x_2$ initially located at the centre of the ring (hatched line), the star denotes the end of this trajectory after 3.8 time-units. None of these trajectories were part of the training or validation datasets. The locations of the particles at each of five timepoints, $t \in \{0.0,0.95,1.9,2.95,3.8\}$, are computed using either the trained ANN (top) or Brutus (bottom) and these solutions are denoted by the bold line. The results from both methods closely match one another and illustrate a complex temporal relationship which underpins the growth in deviations between particle trajectories, owing to a change in the initial position of $x_2$ on the ring.  }
    \label{fig:choas}
\end{figure}

We consider the ability of the ANN to emulate a key characteristic of the chaotic three-body system: a sensitive dependence to initial conditions. We illustrate this in two ways and in each case, we generate new scenarios which are not included in either the training or validation datasets. First, we estimated the Lyapunov exponent across 4000 pairs of randomly generated realizations using the simulation framework described previously. The realizations within each pair differed due to a small random change (of $\delta = 10^{-6}$ in both coordinate axes\footnote{ $\delta = 10^{-6}$ was identified as the minimum distance between a pair of initialisations that allowed for estimation of the Lyapunov exponent and avoided falling below the minimum resolution required to distinguish between a pair of trajectories (owing to the implicit error in the ANN).}) in the initial location of particle $x_2$. The trajectories were computed across two time-units and the first, fifth (median) and ninth deciles of the estimated Lyapunov exponent were (0.72, 1.30, 2.26), indicating some divergence between pairs of initializations. Second, we generated 1000 realizations in which particle $x_2$ was randomly situated somewhere on the circumference of a ring of radius 0.01 centred at (-0.2,0.3) and computed the location of the particle for up to 3.8 time-units, using both the trained ANN and Brutus (Fig. \ref{fig:choas}). Our findings highlight the ability of the ANN to accurately emulate the divergence between nearby trajectories, and closely match the simulations obtained using Brutus (notably outside of the training scenarios).

\begin{figure}
    \centering
    \includegraphics[width=0.45\textwidth]{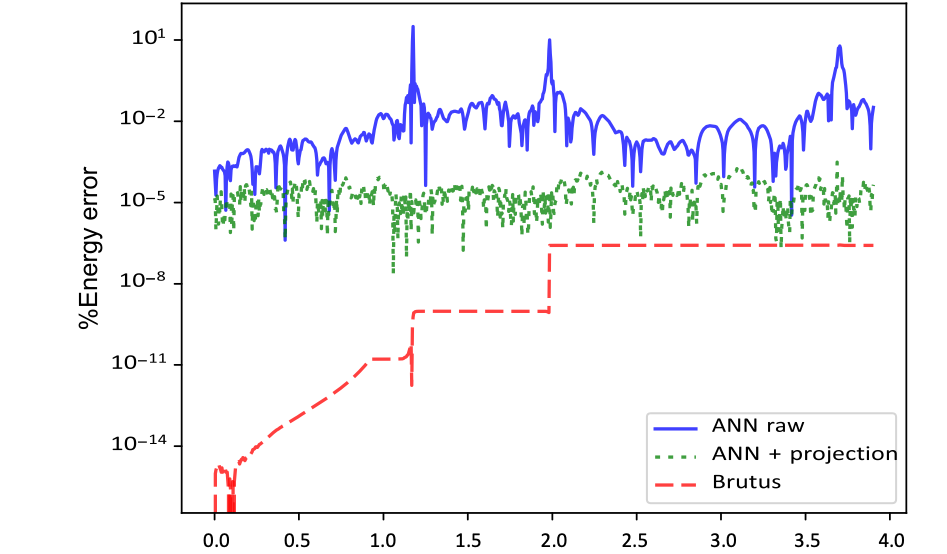}
    \caption{{\bf  Relative energy error. } An example of the relative energy error for a typical simulation. The raw output of the two ANN's typically have errors of around $10^{-2}$, after projecting onto a near by energy surface the error reduces down to order $10^{-5}$.   }
    \label{fig:err}
\end{figure}

We propose that for computationally challenging areas of phase-space, our results support replacing classic few-body numerical integration schemes with deep ANNs. To strengthen this claim further, we assessed the ANN's ability to preserve a conserved quantity, taken as the initial energy in the system, as this is an important measure of the performance of a numerical integration scheme. To do this, we required the velocities of the particles. Our ANN was trained to recover the positions of the particles for a given time, the results from which can be used to estimate the velocities by differentiating the network. Instead, we trained a second ANN to produce the velocity information. A typical example of the relative energy error is shown in Fig. \ref{fig:err}. In general, the errors are of order $10^{-2}$, however, these can spike to $10^1$ during close encounters between the bodies in which case energy is highly sensitive to position. Improvements are achieved by adding a projection layer to the ANN (see Appendix \ref{app:proj}), which projects the phase-space co-ordinates onto the correct energy surface, thereby reducing errors down to around $10^{-5}$. The approach is similar to solving an optimisation problem that aims to find a change in co-ordinates which reduces the energy error whilst also remaining close to the co-ordinates predicted by the ANN.

\section{Discussion}

We have shown that deep artificial neural networks produce fast and accurate solutions to the computationally challenging three-body problem over a fixed time interval. Despite the simplifications in our initial setup, the particle trajectories regularly undergo a series of complex interactions and the ANN captures this behaviour, matching predictions from arbitrarily accurate and precise numerical integrations at a fraction of the computational time cost. For regions of phase-space in which a traditional integrator fails to compute a numerical solution within a pre-specified time-tolerance, it is possible that the trained ANN can be used to provide accurate predictions of the particle's locations away from the present computationally challenging region. These predictions can then be used as input variables to restart the traditional integrator at a future time-point. This idea would see a hybrid numerical integrator developed which combines the traditional integrator with the trained ANN - to calculate particle trajectories local to regions in which the traditional integrator is computationally cumbersome - so that accurate and timely solutions to the three-body problem are obtained over a wider variety of scenarios than is currently achievable.

Three-body interactions, e.g. between a black-hole binary and a single black-hole, can form the main computational bottleneck in simulating the evolution of globular star clusters and galactic nuclei. As these events occur over a fixed time length, during which the three closely interacting bodies can be integrated independently of the other bodies comprising the cluster or nuclei, we have demonstrated, within the parameter space considered, that a trained ANN can be used to rapidly resolve these three-body interactions and therefore help toward more tractable and scalable assessments of large systems. 

With our success in accurately reproducing the results of a chaotic system, we are encouraged that other problems of similar complexity can be addressed effectively by replacing classical differential solvers with machine learning algorithms trained on the underlying physical processes. Our next steps are to expand the dynamic range and relax some of the assumptions adopted, regarding symmetry and mass-equality, in order to construct a network that can solve the general 3-body problem. From there we intend to replace the expensive 3-body solvers in star cluster simulations with the network and study the effect and performance of such a replacement. Eventually, we envision, that network may be trained on richer chaotic problems, such as the 4 and 5-body problem, reducing the computational burden even more.

\section*{Acknowledgements}

It is a pleasure thank Mahala Le May for the illustration of Newton and the machine in Figure \ref{fig:newtonann} and Maxwell Cai for discussions. The calculations ware performed using the LGM-II (NWO grant \# 621.016.701) and the Edinburgh Compute and Data Facility cluster Eddie (http://www.ecdf.ed.ac.uk/). In this work we use the matplotlib  \citep{H2007}, numpy \citep{O2006}, AMUSE \citep{PZat2013,PZMM2018}, Tensorflow \citep{Aat2016}, Brutus \citep{BPZ2015} packages.
PGB acknowledges support from the Leverhulme Trust (Research Project Grant, RPG-2015-408). CNF and PDWK were supported by the MRC (MC\_UU\_00002/7 and MC\_UU\_00002/10). TB acknowledges support from Funda\c{c}\~{a}o para a Ci\^{e}ncia e a Tecnologia (FCT), within project UID/MAT/04106/2019 (CIDMA) and SFRH/BPD/122325/2016. 







\appendix

\section{Tuning and assessing Brutus performance parameters}
\label{app:Brutus}

The Brutus integrator is sensitive to the choice of two tuning parameters: (i) the tolerance parameter ($\epsilon$), which accepts convergence of the Bulirsch-Stoer method \citep{BS1964} and; (ii) the word length ($L_w$), which controls numerical precision. To account for this, interactions with the same initial condition and two choices for the pair {$\epsilon$,$L_w$} were performed 
({$\epsilon=10^{-11},L_w=128$} and {$\epsilon=10^{-10},L_w=88$}). We then calculated the average phase distances $\delta_{a,b}$  between two solutions $a$ and $b$ \citep{M1964}, i.e.

\begin{equation}
\delta_{a,b}^ 2=  \frac{1}{12}  \sum_{i=1}^3 \sum_{j=1}^4 \bigg( q_{\{a,i,j\}} - q_{\{b,i,j\}}  \bigg)^2, \label{eq:delta} 
\end{equation}

to assess sensitivity between the choices over the phase space co-ordinates $q_{\{\cdot ,i,j\}}$ , where $i$ denotes a particle and $j$ its position or velocity. Note $\delta$ is equivalent to the mean squared error of the phase space coordinate. Converged solutions were identified when the average phase distance was $<0.1$ over the duration of a simulation. In over $90\%$ of simulated scenarios we identified converged solutions, exceptions were generally owing to particles initialized near the singularity (Fig. \ref{fig:initialcond}). We also noted sensitivity to the maximum time-step assumed, however we found good resolution of the trajectories when returning results every $2^{-8}$ time units.

\section{Deep artificial neural network}
\label{app:DNN}
Our artificial neural network consisted of 10 densely connected layers with 128 nodes using the $\max{(0,x)}$ (ReLu) activation function, the final output layer used a linear activation function. We considered a variety of ANN architectures. Starting with a network consisting of 5 hidden layers and 32 nodes we systematically increased these values until the ANN accurately captured the complex trajectories of the particles, which we identified as a network containing 10 hidden layers with 128 nodes. We further assessed performance using an ANN with transposed convolution layers (sometimes referred to as a de-convolution layer) which allows for parsimonious projections between a medium and high-dimensional parameter-space. However, performance, as measured by the mean absolute error, was poorer under these networks.  

\section{Projection Layer}
\label{app:proj}
To better preserve a conserved physical quantity, e.g. energy, during the training of a single ANN we introduced a projection layer. The projection layer adjusted the coordinates by minimising the following optimization problem:

\begin{equation}
Er(x,v)^2  + \gamma_1 D_x(x)^2  +  \gamma_2 D_v(v)^2
\end{equation}

where $Er(x,v)$ is the energy error, $D_x$ ($D_v$) is the distance from the initial position (initial velocity) produced by the ANN. Additionally, $\gamma_1$ and $\gamma_2$ are constants which penalise deviation from the initial values of $x$ and $v$, respectively. The optimization problem was solved using the Nelder-Mead method. Instead of this, a training metric, e.g. a fixed multiple of the mean absolute error, can be introduced to bound the error of a prediction.
If a single ANN was trained to predict both position and velocity information for each particle, an alternative strategy would be to introduce a penalty term in the cost function during training (similar to a regularization process).


\bsp	
\label{lastpage}
\end{document}